\documentclass[10pt]{sig-alternate}

\newcommand{\ignore}[1]{}
\usepackage{fancyhdr}
\usepackage[normalem]{ulem}
\usepackage[hyphens]{url}
\usepackage{hyperref}
\usepackage{array}
\usepackage{lipsum}
\usepackage{multirow}
\usepackage{mathtools}
\usepackage{subfigure}
\usepackage{pifont}
\usepackage{caption}
\usepackage{tabularx}
\usepackage{booktabs}
\usepackage{dcolumn}
\usepackage{xcolor}
\usepackage{minibox}
\usepackage{enumitem}
\usepackage{authblk}

\usepackage{microtype}
\usepackage[USenglish]{babel}

\usepackage[letterpaper,left=0.75in,right=0.75in,top=1in,bottom=1in]{geometry}

\title{\vspace{-2ex}Estimating Silent Data Corruption Rates \\ 
Using a Two-Level Model\vspace{-0.3in}}
\author{Siva Kumar Sastry Hari$^1$, 
Paolo Rech$^2$,
Timothy Tsai$^1$, 
Mark Stephenson$^1$,
Arslan Zulfiqar$^1$,
Michael Sullivan$^1$,
Philip Shirvani,
Paul Racunas$^1$,
Joel Emer$^1$,
Stephen W. Keckler$^1$
\\
$^1$NVIDIA Corporation, 
$^2$Federal University of Rio Grande do Sul
}

\begin{document}
\maketitle
\pagestyle{plain}


\begin{abstract}

High-performance and safety-critical system architects must 
	accurately evaluate the application-level silent data corruption (SDC) rates of
processors to soft errors.  
Such an evaluation requires error propagation all the way from
particle strikes on low-level state up to the program output. Existing
approaches that rely on low-level simulations with fault injection cannot
evaluate full applications because of their slow speeds, while
application-level accelerated fault testing in accelerated particle beams is often
impractical.  We present a new two-level methodology for application resilience
evaluation that overcomes these challenges. The proposed approach decomposes 
application failure rate estimation into: (1)~identifying
how particle strikes in low-level unprotected state manifest at the
architecture-level, and (2)~measuring how such architecture-level manifestations
propagate to the program output.  We demonstrate the effectiveness of this
approach on GPU architectures. We also show that using just one of the
two steps can overestimate SDC rates and produce different trends---the
composition of the two is needed for accurate reliability modeling.

\end{abstract}

\section{Introduction}
\label{sec:intro}

Transient hardware faults caused by high-energy particle strikes are of rising
concern for processors deployed in high-performance computing systems
and safety-critical embedded systems.  These transient faults can propagate to
the application level and cause execution failures, also known as Detected
Unrecoverable Errors (DUEs), or worse they can silently corrupt the application output
and lead to Silent Data Cor\-rup\-tion (SDC).

The SDC rate of a system is fundamentally architecture and application dependent.  
Despite this cross-layer complexity, system and software architects 
for high-performance or safety-critical systems 
must ensure that the applications running on their systems achieve acceptable
SDC rates.  Furthermore, software developers also want to gain insight into why their
applications produce certain SDC rates, if these rates are unacceptable, and
how to ensure that their applications are reliable with minimal impact on
performance and power.

These objectives necessitate a fast and accurate resilience evaluation technique.
Quantifying how particle strikes in unprotected low-level state propagate to the application output requires
a detailed design such that transient bit-flips in unprotected structures
can accurately be injected or modeled and a fast evaluation framework that
propagates such low-level errors all the way to the program output, and not just to
the output of a chip or a kernel.  Meeting these conflicting requirements makes application
SDC rate estimation challenging. 

Existing approaches are insufficient to investigate app\-li\-ca\-tion-le\-vel
resilience at the level of detail required to enable software architects to
quantify the SDC rates of their applications and to develop insights into what
makes their applications vulnerable---insights that are key for
developing low-cost resilience solutions. We categorize existing approaches
into five following groups. Table~\ref{tab:comparison} compares
these approaches based on their speed, accuracy, complexity, and visibility into
the application state corruption.

\newcommand\introcolwidth{1.6cm}

\begin{table*}[htbp]
\begin{center}
    \scriptsize 
    \caption{A comparison of the various techniques for deriving application-level SDC rates.}
    \vspace{-0.1in}
    \label{tab:comparison}
    \begin{tabular}{| p{4cm} | p{1.2cm} | p{1.2cm} | p{1.2cm} | p{1.4cm} | p{1.5cm} | p{1.3cm} | p{1.2cm} |}
    \hline

    & \multicolumn{2}{c|}{Low-level fault simulation} & \multirow{2}{1.4cm}{Higher-level fault injection} & \multirow{2}{1.4cm}{Hierarchical fault simulations} & \multirow{2}{1.4cm}{Microarchi\-tecture-level ACE analysis} & \multirow{2}{1.4cm}{Accelerated fault testing} & \multirow{2}{1.4cm}{Our approach} \\ \cline{2-3}
    & 	& & & & & & \\ 
    & 	RTL & FPGA-based & & & & & \\ \hline

    Captures application-level error propagation & 
    \textcolor{red}{No} & Yes 	& Yes & Yes & Somewhat & Yes & Yes 	\\ \hline

    Visibility into application state corruption & 
    \textcolor{red}{Low} & \textcolor{red}{Moderate} & High & High & \textcolor{red}{Moderate} & \textcolor{red}{Low} & High \\ \hline

    Fault model accuracy & 
    High & High & \textcolor{red}{Low} & High & \textcolor{red}{Depends on the model} & High & High \\ \hline

    Speed & 
    \textcolor{red}{Slow} & \textcolor{red}{Medium} & Fast & \textcolor{red}{Medium} & \textcolor{red}{Medium} & Fast  & Fast \\ \hline

    Implementation complexity & 
    \textcolor{red}{Medium} & \textcolor{red}{High} & Low & \textcolor{red}{High} & \textcolor{red}{High} & Low & Low \\ \hline

    \end{tabular}
\end{center}
\vspace{-0.2in}
\end{table*}

\begin{itemize}[noitemsep, leftmargin=*]

\item {\bf Low-level simulation-based fault injection} uses either an
  RTL-level or microarchitecture-level simulator to inject faults 
  by flipping bits 
  in low-level state~\cite{SaggeseMICRO2005,Maniatakos2011}. The accuracy
  of this approach is limited by the fidelity of its model---RTL is very
  accurate, microarchitecture simulators less so---and it can only 
  study the impact of faults at the architecture- or chip-level due to
  slow simulation speed. Since error manifestations at these levels may not
  propagate to the application output, full-application analysis is impractical 
  with this approach. FPGA-based
  simulations~\cite{Cho2013DAC, ebrahimi_fast_2014, raghuraman_2014} are faster 
  but they suffer from high implementation complexity, limited visibility into
  application-level error propagation, and limited availability of FPGAs that
  can fit chip designs that contain 10s of billions of transistors. 

\item {\bf Higher-level simulation-based fault injection} either injects
  faults at the architecture-level or into compiler-level intermediate
  representations~\cite{SASSIFI_ISPASS,fang2014GPU-Qin,Wei2014-LLFI,li2012classifying,feng2010shoestring}.
  This approach is fast and provides visibility into application corruption.
  The error modeling accuracy, however, depends entirely on the chosen model,
  and it is often limited to injecting single bit-flips uniformly at the
  instruction-level. 

\item {\bf Hierarchical fault simulations} integrate low- and higher-level
  simulators such that only the required details are modeled at the low-level
  to accurately inject the fault.  Once the error manifests at the
  higher-level, the faster simulator takes over~\cite{Choi1989FOCUS, KalbarczykTSE1999,
  LiHPCA2009}.  While this approach is much faster than low-level simulations,
  the integration complexity makes it impractical for large systems.
 
\item {\bf Microarchitecture-level analysis} attempts to identify
  which values in microarchitecture structures can possibly affect correct
  program execution (also known as ACE
  analysis)~\cite{MukherjeeMICRO2003,BiswasISCA2005,li_understanding_2008,
  RaaschMICRO2015}.  This analysis only covers structures whose occupancy can
  be reasoned about (e.g., large SRAM structures). As these approaches
  typically rely on simulation to characterize occupancy, examining full
  applications is prohibitively expensive.

\item {\bf Accelerated fault testing} subjects an existing chip to a
  high-energy particle beam and measures the observed chip
  error and failure rates~\cite{ziegler_accelerated_1996, Oliveira2015TC}. While
  this approach simulates the long-term behavior of computing systems and can
  be performed on full applications at speed, it does not allow for controlled
  injection or the detailed observability of errors.  It also does not provide 
  any insight into the vulnerability of low-level structures. 

\end{itemize}

To enable fast, accurate, and flexible applicat\-ion-level error analysis,
we propose a novel two-level technique that decomposes application failure rate
estimation into two decoupled components: (1)~identifying and modeling how
particle strikes in low-level unprotected state manifest at the
architecture-level, and (2)~measuring how such architectural manifestations
propagate to the program output. In this paper, we develop tools that enable
this approach to be applied to the NVIDIA GPUs\@.

For the first step, we employ accelerated neutron beam testing 
using carefully crafted test programs on existing silicon as this methodology 
offers realistic fault models for accurate failure rate estimates. 
Each test program repeatedly executes a specific instruction type
and captures architecture-level error manifestation rate from the particle
strikes in unprotected structures (latches, flip-flops, and SRAMs) of the
systolic instruction execution pipelines. For throughput oriented processors
such as GPUs, such structures are expected to be the biggest contributor to
SDCs (our results also support this expectation)~\cite{Jeon2012WarpedDMR, Snir2014}.  While beam experiments are
subject to statistical uncertainty, they eliminate the modeling inaccuracies
that arise in simulation-based approaches.  Capturing architecture-level error
manifestations after every instruction is challenging as the error checking 
and recording logic after every instruction can be prohibitive. We address this 
challenge and present methods to prepare tests for different instruction types.
We call this step {\it Implementation-level Propagation Analysis (IPA)}.

The second step employs fast error injection into the architectural state of a
program running on existing hardware and measures how these errors propagate to the
program output.  This propagation is a function of the program itself rather
than the microarchitecture it runs on. For our GPU case study, we employed 
SASSIFI that instruments a CUDA program with code that can flip
bits in low-level GPU assembly instruction (SASS) outputs~\cite{SASSIFI_ISPASS}.  
As SASSIFI runs directly on the GPU, it is orders of magnitude faster than
detailed simulators and can use unmodified applications.
We call this step {\it Architecture-level Propagation Analysis (APA)}.

In the final step, we estimate an application's SDC rate by combining the
results from IPA and APA along with application- and device-specific
performance metrics. The following are the contributions of this paper. 

\begin{itemize}[noitemsep, leftmargin=*]

	\item A two-level methodology to estimate SDC rates that is
		faster than low-level simulation-based ap\-proaches, more
		accurate than higher-level simulations, and more flexible
		than accelerated application fault testing. This approach
		requires performing IPA experiments just once per GPU
		generation and fast APA experiments per application to derive
		application SDC rate estimates on a target GPU.

	\item We quantify how particle strikes in low-level unprotected
		structures of state-of-the-art GPUs manifest at the
		archi\-tecture-level using accelerated beam experiments. We
		address the challenge of capturing architecture-level error
		manifestations after every SASS instruction with minimal
		error checking and recording logic (by introducing it only
		after a long sequence of target SASS instruction). 

	\item We estimate SDC rates of all workloads from the Rodinia benchmark
		suite~\cite{Rodinia} and two DOE mini-apps (CoMD~\cite{CoMD}
		and Lulesh~\cite{Lulesh}).  We compare these estimates to
		direct beam test results for a selected set of workloads.  

	\item We present insights into how IPA- or APA-only approaches can
		provide inaccurate SDC rates and trends. They overestimate SDC
		rates and show different trends---the composition of the two is
		needed for accurate reliability modeling.  

\end{itemize}

\section{Overview}
\label{sec:motivation}

\begin{figure}[tbp]
    \centering
    \includegraphics[width=0.47\textwidth]{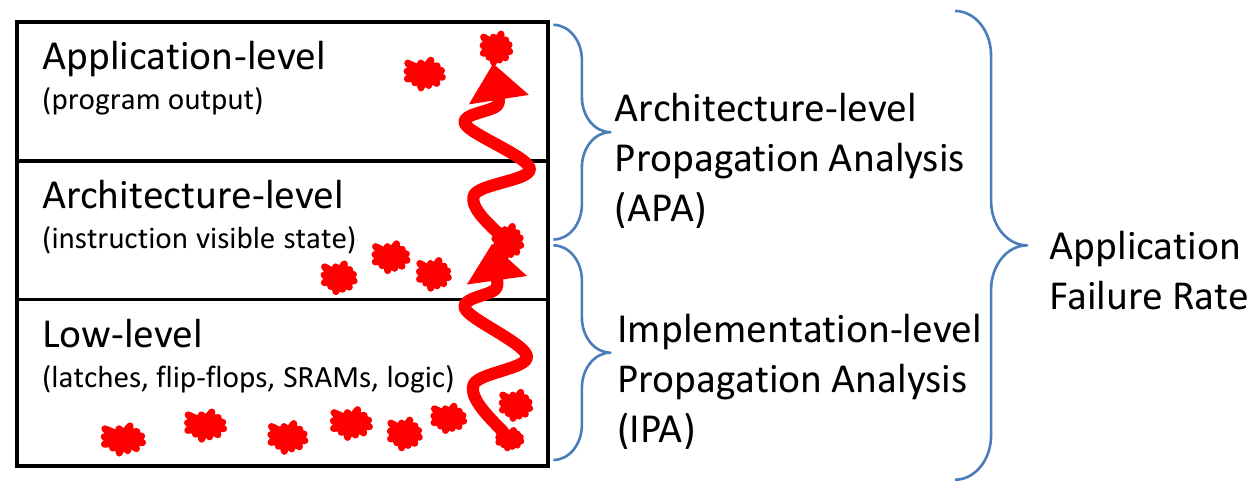}
    \vspace{-0.1in}
    \caption{Composing low-level and architecture-level error propagation
    analyses to estimate application-level failure rate.  The red
    symbols represent faults/errors at different levels of the design
    hierarchy.}
    \vspace{-0.1in}
    \label{fig:avf-analysis}
\end{figure}

Considering the need for a fast, accurate, and flexible application resilience
analysis and the challenges associated with prior solutions (as discussed above
and in Table~\ref{tab:comparison}), we propose a two-level application failure
rate evaluation methodology that accounts for both low-level and program-level
error propagation.  
The first step, IPA, evaluates how particle strikes in
unprotected structures manifest as changes to the architecture-level state.  
The second step, APA, evaluates how such manifestations propagate to the
program output.
As a final step, we combine the results from these two steps along with simple
application- and device-specific metrics to estimate application SDC rate. This
rate is expressed as Failures In Time (FIT), where one FIT equals one failure in a
billion hours.  Figure~\ref{fig:avf-analysis} summarizes our approach.  While
this approach can be applied to any processor, we demonstrate it on 
NVIDIA GPUs\@.

\subsection{Error Model to Bridge IPA and APA} 

\begin{figure}[bp]
    \centering
    \vspace{-0.1in}
    \includegraphics[width=0.48\textwidth]{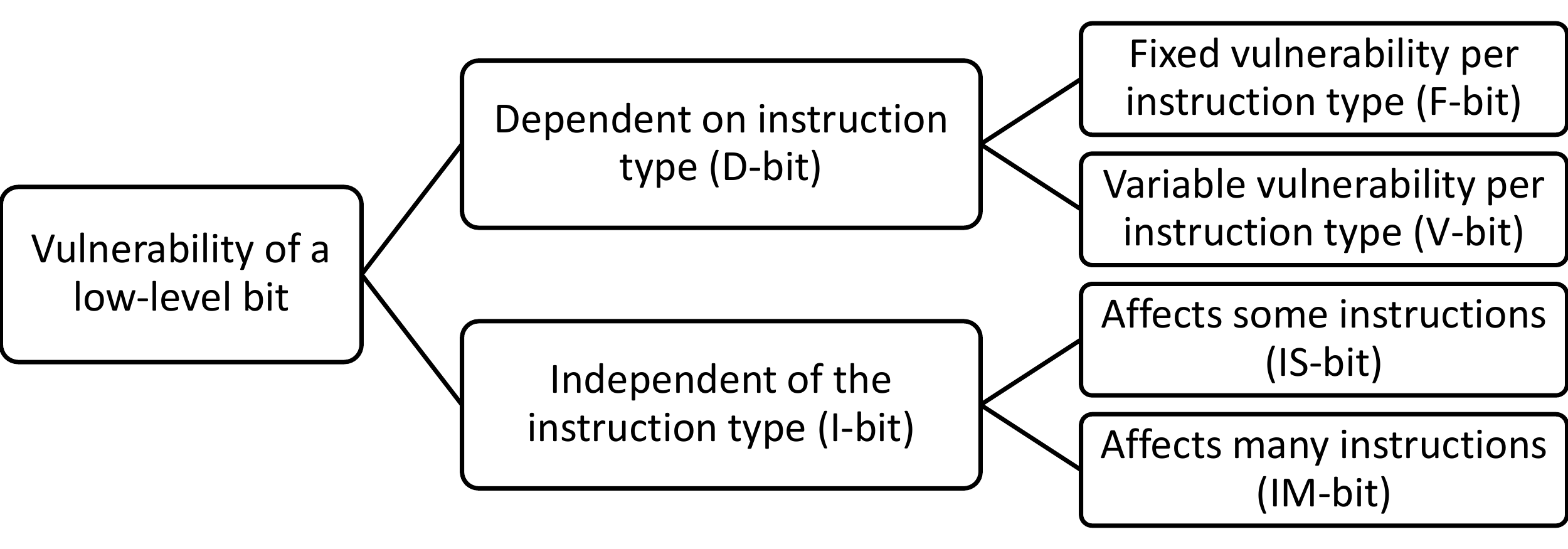}
    \vspace{-0.2in}
    \caption{Categorizing low-level bits.}
    \label{fig:bit-categorization}
\end{figure}

We bridge IPA and APA with an architecture-level error model that captures
bit-flips in low-level unprotected structures as bit-flips in assembly
(SASS) instruction outputs. The bit-flips in low-level state can manifest in
different ways at the architecture-level and the manifestation rate 
depends on which low-level bit is flipped.  Figure~\ref{fig:bit-categorization}
summarizes how we categorize low-level bits.

The vulnerability of a low-level bit is either dependent or independent of
the instruction type that is executing. We call the bits whose vulnerability
depends on the instruction that is executing as D-bits and the ones that do not
as I-bits. For example, bits in ALU are categorized as D-bits and bits in
structures that mediate communication between CPU and GPU are categorized as
I-bits. 

{\bf D-bits: }
Vulnerability of a D-bit can either be fixed or variable per instruction type.
We call the D-bits whose vulnerability is fixed per instruction type as {\it F-bits}.
FIT rate of these bits scales linearly with instruction issue rate.  Examples
of F-bits include flip-flops and SRAM bits in buffers in systolic pipeline
stages.  Since GPUs consist of simple in-order pipelines, bit-flips in F-bits
mostly manifest as bit-flips in the destination register of one instruction in
one thread.  

We call the D-bits whose vulnerability varies with different
microarchitecture-level buffer occupancies per instruction type as {\it V-bits}.  FIT
rate of these bits scales with instruction issue rate as well as other
performance metrics.  Examples of V-bits include unprotected SRAM bits in data
caches, load/store buffers, and DRAM buffers.  Since data can reside in these
structures for a variable amount of time, the vulnerability of such bits will
also vary. 

Bit-flips in V-bits can manifest as DUEs or bit-flips in destination
registers of instructions.  For example, a bit-flip in an unprotected load
buffer entry can only corrupt one instruction's output in just one thread.  The
size of the corruption (number of bit-flips) depends on whether the data or 
address bit was flipped.  A bit-flip in an unprotected buffer between L2 and L1
data caches while the data is being transferred from L2 to L1 cache can
manifest as a bit-flip in a cache line in L1 cache. 
If the cache line has only one reuse, the corruption
may affect the destination registers of only one instruction. If the line has
several uses, destination registers of multiple instructions in multiple
threads in multiple warps can be corrupted. 

{\bf I-bits: }
We categorize I-bits further based on the number of instructions that could be
corrupted by the underlying bit-flip. We call the I-bits that corrupt only some
instructions as {\it IS-bits} and the bits that corrupt many instructions as
{\it IM-bits}.  For example, a bit-flip in an unprotected instruction buffer in the
front-end of the pipeline can corrupt the destination registers of all the
threads in one warp (due to the SIMT nature of the GPUs). If the corrupted
instruction is used by multiple warps then it can corrupt multiple warps.  The
number of instructions corrupted by this bit are, however, limited. So we call
it an IS-bit. 
We categorize the bits that are not directly used for a specific instruction's
execution, but are needed for correct functioning of the GPU as IM-bits. One
example of such bits is a bit in a structure used to mediate communication
between CPU and GPU\@. A bit-flip in it can corrupt many data words and
manifest in many instructions during execution. 

In this paper, we focus on modeling the impact of bit-flips in the F-bits.  Our
IPA results, as discussed later in Section~\ref{sec:ipa-results} indicate that
the F-bits are the primary sources of the SDCs.  We also show the potential
impact of bit-flips in the V-, IS-, and IM-bits.  

We model architectural manifestations from bit-flips in F-bits using a 
per-instruction-type architecture-level model. For example, a particle strike in a
pipeline latch while an integer add instruction is executing can propagate to
the destination (output) register value and corrupt just one bit location. We
capture this as a single bit-flip in the destination register of the integer
add instruction.  We also capture the manifestations from a single event per
instruction type as (1)~multiple bit-flips in the destination register
of one instruction, and (2)~single and multiple bit-flips in the destination
registers of one instruction that spans multiple threads in a warp.  These
error outcomes stem from our observations from IPA experiments, as discussed in
Section~\ref{sec:ipa-results}. We use this categorization later to conduct APA
experiments.

\subsection{IPA}

IPA can be performed in multiple ways to capture such architecture-level
manifestations. Accelerated high-energy neutron beam testing using
state-of-the-art GPUs, low-level (RTL) error injection based studies, and ACE
analysis using detailed microarchitecture-level simulators are three such ways.
While the latter two approaches can be performed in pre-silicon stages, the
availability of the detailed simulation infrastructure, simulation speeds, or
design complexity often become the limiting factors.  We chose the first
approach because (1)~the phenomenon that induces errors is closest to the
natural occurrences of soft errors,  (2)~we do not have modeling inaccuracies
that simulation based approaches suffer, and (3)~we do not require detailed
low-level simulators.  For the beam testing, we designed workloads that
repeatedly execute a specific instruction per workload and capture
manifestations in the outputs of each instruction. The output of this analysis
is a set of error propagation rates for different manifestations (e.g., single-
or double-bit flips in one destination register) for each instruction type. The
manifestation may also be a program crash or hang. 

\subsection{APA}

APA can be performed in a couple of ways such as dynamic program-level
liveness analysis or software-based instruction-level error injection.  
We choose the second approach primarily because it can
precisely measure program-level error propagation (across multiple kernels in
case of GPU programs) to the output.  For this step, we employ a fast
instruction-level error injection tool called SASSIFI that can
inject into workloads running at-scale directly on the silicon~\cite{SASSIFI_ISPASS}. For all
the error manifestations observed in IPA, we perform statistical error
injection campaigns per instruction type for all our workloads and obtain
application-output level propagation probabilities. 

\subsection{Composing IPA and APA to Estimate SDC Rate}

We estimate application-level failure rates by composing the
results from the IPA and APA with per-application dynamic instruction
distributions and device-specific performance metrics.  
We compute FIT rates for both SDCs and DUEs, but focus on SDCs. 
We use the Equation~\ref{eqSDC} to compute the per-application SDC FIT rate for
F-bits. 
This equation essentially multiplies the rate of manifestation of an
incorrect architecture state (IAS) per instruction type, obtained during IPA,
with the probability of a SDC given such architecture-level
manifestation obtained during APA.  It factors the instruction distribution
of the application. This result is then scaled with the application's
instruction issue rate on the specific device for F-bits to account to obtain
the application FIT rate (lower the issue rate, the lower the SDC rate if other
parameters do not change). Per instruction type issue rate might result in a
more accurate estimate, but is hard to obtain on silicon.  
This formulation can be used for V- and I-bits as well, but the scaling factor
will depend on other performance metrics. 

\vspace{-0.1in}
\begin{align}
	\label{eqSDC}
	\begin{split}
	FIT_{SDC} =  
	( \displaystyle\sum_{n=1}^{N} f_n \times 
	( \displaystyle\sum_{m=1}^{M} pIAS_{nm} \times pSDC_{nm} ) )
	\times s 
	\end{split}
\end{align}
\vspace{-0.2in}

\noindent $where$ \\
{\em N = number of instruction types in an application} \\
$f_n$ {\em = fraction of application instructions of type n} \\
{\em M = number of architectural error manifestations, including bit flip patterns} \\
$pIAS_{nm}$ {\em = rate of incorrect architectural state for a specific manifestation m for instruction type n} \\
$pSDC_{nm}$ {\em = probability that the manifestation m in instruction type n will result in an application-level SDC} \\
{\em s = application and device specific scaling factor (issueIPC in this study)} \\

\vspace{-0.05in}
We obtain $pIAS_{nm}$ values during IPA and $pSDC_{nm}$ during APA. For the IAS
that are crashes and hangs, $pSDC_{nm} = 0$.  We obtain $f_n$ and $s$
through dynamic application profiling.

\section{IPA Methodology}
\label{sec:ipa}

The goal of IPA is to address two key questions. (1)~How do low-level
errors in state-of-the-art GPUs manifest at the outputs of the executing
instructions?  For example, do particle strikes in unprotected flip-flops,
latches, and SRAMs propagate as single or multiple bit flips in destination
registers? Do they corrupt single or multiple instructions? How many threads do
they corrupt?  (2)~What are the rates for different instruction-level error
manifestations? 

Focusing on F-bits, our approach is to capture archi\-tecture-level
manifestations after each instruction while running microbenchmarks on GPUs in
an accelerated high-energy neutron beam.  We develop a microbenchmark suite
targeting seven commonly used SASS instructions. These instructions constitute
nearly half of the dynamic instruction count in our workloads (see
Section~\ref{sec:apa-results} for details). We expect the vulnerability
from F-bits to be significant for these instructions. Each microbenchmark
captures the architecture-level manifestations of particle strikes in low-level
state after virtually every instruction execution. We manually attribute the
observed events to F-, V-, IS-, or IM-bits.  This approach provides us an estimate of
the relative contribution of different types of bits to SDCs. It allows us to
measure the rates for different instruction output manifestations (e.g.,
single or multiple bit flips or random values in one or multiple instructions
in one or multiple threads) for the different SASS instructions. 

\subsection{Microbenchmark Suite to Quantify Archi\-tecture-level Error Manifestations}

The challenging part in developing such microbenchmarks is to capture all
instruction-level manifestations while ensuring that the checking and recording code does
not significantly perturb normal execution sequence of the instruction under test. The
key insight we use to overcome this challenge for arithmetic instructions is to
accumulate values produced
by instructions into an accumulator register.  We preserve the destination
registers of all the instructions until we compare the accumulator's value to a
predetermined constant to check for error manifestations.  If the
comparison fails then we write all the register values to host pinned memory.
This allows us to identify which instruction(s) observe the manifestation and
which propagate the manifested corruption.  Since the number of available
registers per thread is limited, we need to perform the check after a certain
number of back-to-back arithmetic operations. Based on the per-thread register
availability, this approach allows us to keep the checking code and recording
code to $<$10\% for our microbenchmarks. 

We test with ECC on, which means the register file, L1 and L2 caches, shared
memory, and DRAM are protected from single-bit flips.  
We record but ignore uncorrected ECC errors in this study. We develop seven
microbenchmarks that target the seven commonly used SASS instructions based on
dynamic instruction profile of the workloads from the Rodinia benchmark suite:
IADD, FADD, IMAD, FFMA, LDS, ISETP, and BRA~\cite{NVIDIA:ISA}.  This mix
contains integer, floating point, shared memory load, and control instructions.
For each target instruction, we write a CUDA kernel such that the targeted
instruction executes repeatedly to dominate the total runtime.  These
programs have almost no control and memory divergence.  For the most part, the
threads operate completely out of the register file and do not use the memory
subsystem (except for the I-caches).

\begin{figure}[tbp]
    \centering
    \includegraphics[width=0.5\textwidth]{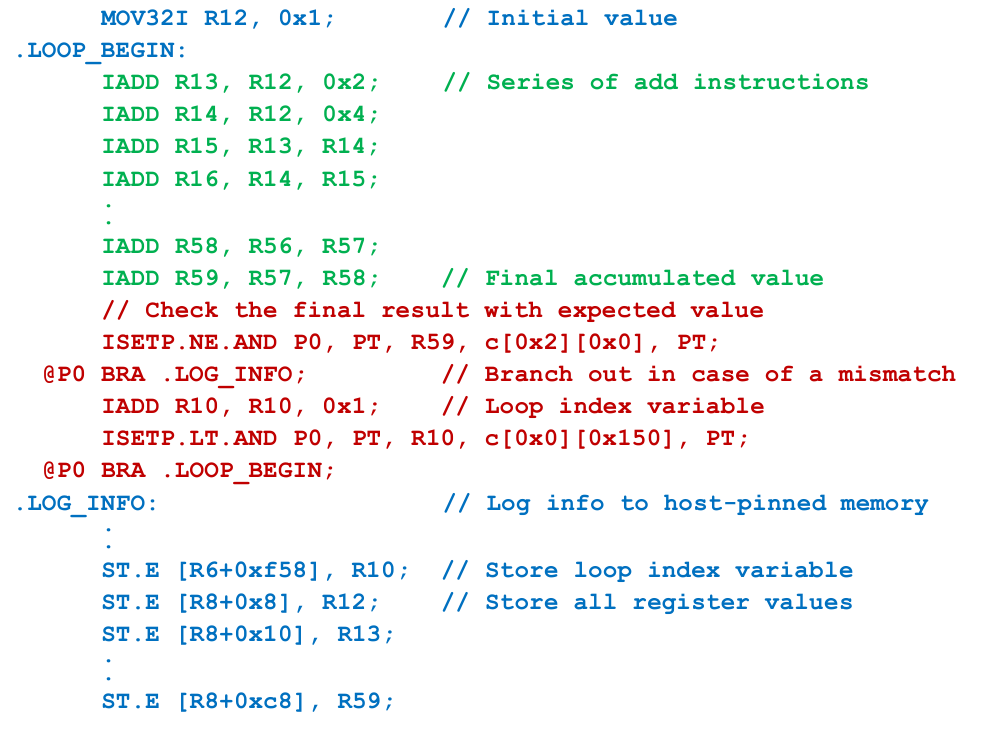}
    \vspace{-0.25in}
    \caption{The main section of the microbenchmark that tests how particle
    strikes in IADD manifest at the architecture-level.}
    \label{fig:iadd_example}
    \vspace{-0.1in}
\end{figure}

{\bf Arithmetic Instructions: }
For the IADD kernel, we first initialize the register content and then execute
a long sequence of IADD instructions that executes a Fibonacci series. After the series, which is
about 45 instructions long in our setup, we compare the final value with the
expected value.  If a mismatch is detected,  we store all the register values
(outputs of each of the instructions that are part of the Fibonacci series) to
host memory to be logged by the host for post-processing.
Figure~\ref{fig:iadd_example} shows the main section of this program.  We
prepared programs for the other arithmetic instructions (FADD, IMAD, and FFMA)
using a similar approach, i.e., accumulating values from a sequence of
instructions to reduce interference from the checking and recording code. 
This approach allows us to map the every manifestation back to the instruction
where it originated and how it manifested (in one or multiple instructions).

{\bf Compare Instruction: }
An ISETP instruction compares the input operand registers and writes the
one-bit result into a predicate register. There are seven one-bit predicate
registers in the Kepler ISA (our target GPU is Kepler-based)~\cite{NVIDIA:ISA}.
Our ISETP kernel also contains a long loop that executes a sequence of ISETP
instructions with a few other instructions. We hand-modify the body of this
loop such that we execute seven back-to-back ISETP instructions (that write to
different predicate registers) followed by a P2R instruction;\footnote{We use
non-publicly available tools to generate and assemble this hand-modified SASS
code.} this moves all seven predicate bits (along with condition codes) into a
general purpose register.  We accumulate the value of these general purpose
registers, similar to our IADD kernel, and move them to host memory if the
accumulated value does not match the expected value.  This provides us the
ability to map every manifestation back to the instruction where the error
would have originated. We can also identify if multiple instructions in one or
multiple threads manifest at a similar time. Through this approach we may
attribute an error in the P2R instruction as an error in the ISETP instruction,
which is one of the limitations.

{\bf Branch Instruction: }
For the BRA program, we write a chain of branch instructions such that the
target of each branch instruction is another branch instruction. We separate
these instructions by tens of instructions and insert filler branch
instructions that jump to a record routine and terminate program.  These filler
branch instructions do not execute on a fault-free run. We execute these
chained branch instruction in a loop that terminates after a certain number of
iterations.  The record routine logs the loop index variable and allows us to
detect whether the loop terminated early.  An error in the executing branch
instruction can result in the following three events. (1) Control transfer to a
filler instruction: The designed program will terminate early (without
completing predetermined loop iterations) and record the loop index variable
for post-processing. (2) Control transfer to outside the program: Program
crashes in these cases. (3) Control transfers to non-filler instructions: Some
of these errors can go undetected and we reduce this probability by limiting
non-filler instructions. 

{\bf Shared Memory Load Instruction: }
For the loads from shared memory (LDS), we first write known values
to the all the shared memory locations. We load them to different registers in
a loop iteration. We accumulate the values into an accumulator, using a similar
concept as in IADD program above, and record the register content for later
inspection if the accumulator does not match its expected value.  In this
program we can distinguish whether the IADD or the LDS instruction noticed the
error as the IADD instructions should only corrupt the accumulator not the
individual register. The data in shared memory is protected by ECC from direct
particle strikes.

\subsection{Beam Experiments}
\label{sec:ipa:beam}

\begin{figure}[tbp]
    \centering
    \includegraphics[width=0.47\textwidth]{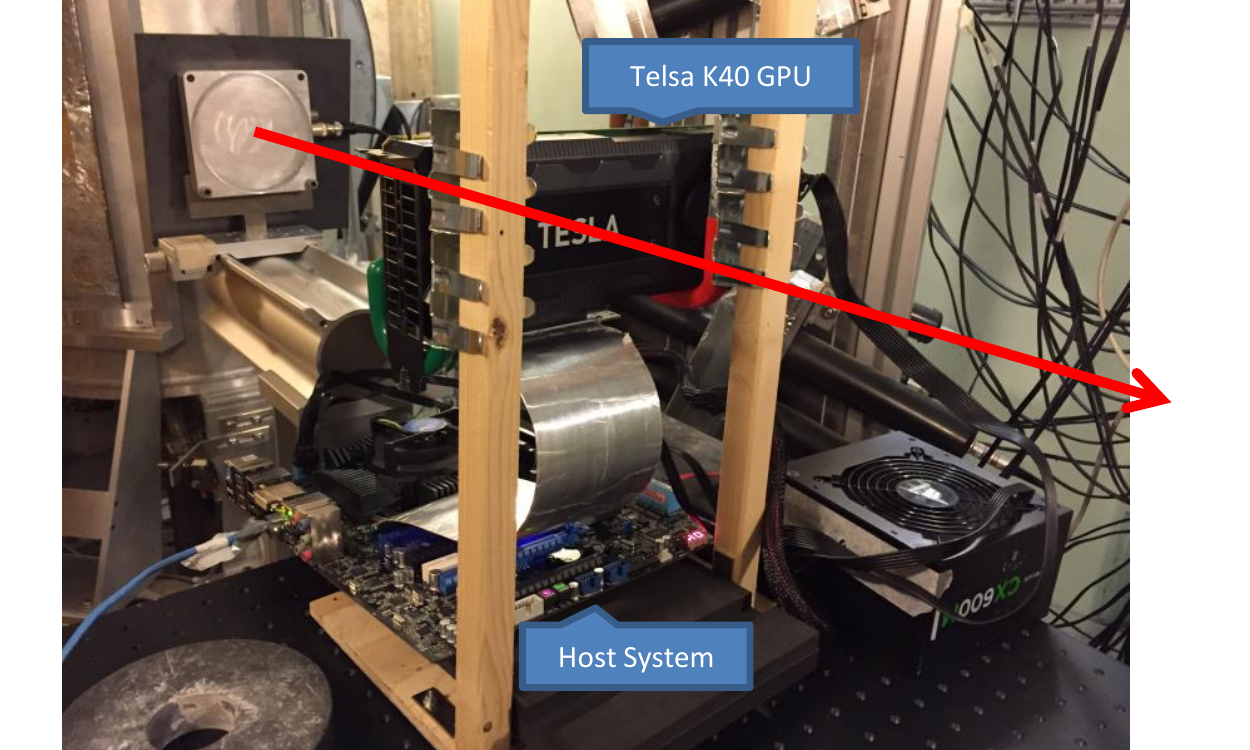}
    \vspace{-0.05in}
    \caption{A picture of the experimental setup at the ISIS beam testing
    facility.} 
    \label{fig:isis_setup}
\vspace{-0.15in}
\end{figure}

We conducted the beam experiments to quantify architectural manifestations for
the arithmetic and compare instructions (IADD, IMAD, FADD, FFMA, and ISETP) at
the ISIS facility near Oxford, United Kingdom and for the branch and shared
memory load instructions (BRA and LDS) at the LANSCE facility at Los Alamos,
New Mexico using NVIDIA K40 Tesla GPUs~\cite{TeslaK40}.
Figure~\ref{fig:isis_setup} shows the K40 board and the accompanying hardware
in the beam experiment room on the first day of testing at ISIS.  The neutron
flux at these facilities is 4-7 orders of magnitude higher than the flux at sea
level.  We adjusted the microbenchmark runtimes (using loop iteration counts)
such that the probability of observing only one architectural-visible error
(crash or value corruption) due to a particle strike is high.  This allow us to
scale accelerated beam test results to realistic environments.  We calibrate
the FIT rates obtained from different beam testing facilities based on a common
test with known FIT rate. 

The K40 board includes a Kepler architecture based GK110b GPU
chip and 12 GB GDDR5~\cite{KeplerArch}. We irradiated just the GPU chip. The
chip is fabricated using 28nm planar bulk technology from TSMC and includes 15
Streaming Multiprocessors (SMs), up to 2048 threads/SM, 30 Mbit total register
file (RF), 7.86 Mbit total L1 cache/Shared memory, 12 Mbit L2 cache.  The
register files, shared memory, caches, and DRAM are ECC protected and we run
with ECC on.

We compute the FIT rate per architectural manifestation per instruction using fluence
(total number of high-energy neutrons), event counts, and New York City-level
flux (13 neutrons/cm$^2$/hour~\cite{JEDEC}). In this paper, we only
show relative FIT rates. 
We conducted tests over several days with an effective beam time of
over fifteen hours with a single GPU for these tests. The total beam usage
time was higher, considering the setup and down time.
The 95\% confidence interval error bars for the
FIT rate per microbenchmark are 
	[42\%, -29\%] for IADD,  
	[49\%, -33\%] for FADD,  
	[42\%, -29\%] for IMAD,
	[38\%, -27\%] for FFMA, 
	[261\%, -69\%] for ISETP,
	[40\%, -28\%] for BRA, and  
	[58\%, -36\%] for LDS.
The error bars are not tight because we used only one GPU for this campaign
(parallel experiments with multiple GPUs can lower the error bars
significantly). We, however, observed a variety of instruction-level manifestations
including single, double, 3+ bit flips in one destination register
in a single thread, double and 3+ bit flips in one destination register across
the threads in a warp for different instruction types. The results are shown in
Section~\ref{sec:results}.

\subsection{Event Categorization and Attribution} 
\label{sec:ipa:attribution}

We run our microbenchmarks back-to-back while logging state (using heartbeats)
of each run at various points in the program and recording any errors that are
reported by the program or system.  If a program takes more than 3x its
expected runtime, we terminate it and call the event a hang. If the system
becomes unresponsive, we restart it. On every failure (crash or hang), we reset
the GPU (using nvidia-smi tool~\cite{nvidia-smi}) before continuing with the
next run.  During program execution, we record status messages before and after
the GPU kernel and flush the contents to a file, which help us to identify 
the failures that occur during the GPU kernel execution. 

If a failure occurs before the kernel starts, we attribute it to a bit-flip in
an IM-bit. If a hang is detected during a kernel execution, we have visibility
in diagnosing the root cause, and cannot attribute the event to a low-level bit
type. If
the kernel or program crashes during kernel execution, we inspect the
error code as reported by the CUDA error handler or the Linux kernel log (which
we collect during LDS and BRA experiments) to attribute the error to a
particular bit type. We attribute the illegal-instruction error to IS-bits
because we hypothesize that a bit-flip in an unprotected buffer while
fetching the instruction results in such a manifestation. We attribute the
out-of-range errors and the stack errors (which occurred only during the LDS
and BRA experiments, respectively) to F-bits. 

On a program completion (without a crash or hang), we inspect the recorded values and 
compare them to the expected values. If all the values match, we assume that
either the loop control or checking logic in the microbenchmark is affected
(not the targeted instructions), which triggered the recording routine. We
ignore such events. On a value
mismatch, we analyze the corruption and categorize it as (1)~single and 
(2)~double bit-flips in one instruction, (3)~a random value error in one instruction,
(4)~a random value error in same instructions in two threads, (5)~double
bit-flips in the same instruction across all threads in one warp, (6)~a random value
error in the same instruction across all threads in one warp, (7)~a zero value in
the same instruction across all threads in one warp, or (8)~a random value error
in two instructions across all threads in two warps. As events 1-4 corrupt just
one instruction's output in just one or two threads, we hypothesize that the
source of corruption to be a bit-flip in a pipeline element. Hence, we
attribute events 1-4 to
F-bits. We hypothesize that events 5-8 are caused by bit-flips in the
instruction(s) because they affect all of the threads in a warp in a same way
(GPUs are SIMT in nature). Event 8 might be due to a multi-bit error that spans
multiple rows in an unprotected buffer. Based on this hypothesis, we attribute
events 5-8 to IS-bits.

\section{APA Methodology}
\label{sec:apa}

The second step in estimating an application's SDC FIT rate is to quantify how
instruction output-level errors propagate to the application output.
We used SASSIFI in this step to inject transient errors into ISA-visible state
such as general purpose registers, predicate registers, and condition
codes~\cite{SASSIFI_ISPASS}. SASSIFI leverages the SASSI~\cite{SASSI_ISCA},
which provides the ability to instrument instructions in the SASS code.  
SASSIFI injects instrumentation before and after the instructions to identify
injection sites and to inject the error by modifying destination register
values, respectively. 

SASSIFI operates in three main steps: (1)~profiling the application to identify
possible error injection sites; (2)~statistically selecting error injection
sites for each error model; and (3)~injecting errors into applications based on
the selected error model and monitoring the resultant behavior.  The error
model is defined by the IPA\@. 

{\bf Profiling to Identify Error Injection Sites:}
In the profiling step, SASSIFI collects names of static kernels and the number
of times they execute during a fault-free application run to to identify the
error injection space.  It also collects the number of dynamic instructions of
each opcode and instruction type per kernel invocation.

Since we have IPA results for seven instructions, we group instructions that we
expect will propagate bit-flips in low-level state to the architecture-level
similarly to the ones we tested.  We examine the Kepler ISA and mark the
instructions that operate solely on architectural registers to be similar to
either IADD, FADD, IMAD, FFMA, or ISETP\@. We mark several control instructions
to be similar to BRA\@.  Since instructions in a group are expected to exercise
similar low-level pipeline structures, their inherent vulnerabilities should be
strongly correlated.

As an example of these groupings, we place integer maximum/minimum, shift
operations, logical operations, and float-to-integer instructions in the same
group as IADD.  In this study, we group single and double precision
floating-point operations together. Placing them in different groups with
associated IPA experiments will improve the accuracy further. 

{\bf Statistically Selecting Error Injection Sites:}
For an architecture-level error model (from IPA) and the respective instruction
group, SASSIFI statistically selects hundreds of dynamic instructions among all
the dynamic kernel executions as injection sites. An error injection site is a
tuple consisting of the instruction group ID, architecture-level error model,
static kernel name, dynamic kernel invocation ID, dynamic instruction count,
seed to randomly select a destination register, and seed to select the error
for injection based on the error model (e.g., location of the bit flip for a
single-bit flip model).  Error bars are under 3\% at 95\% confidence intervals
for our measured SDC probability per error model and instruction type. 

{\bf Error Injections Runs:} 
We use the {\it Instruction Output Value (IOV)} mode in SASSIFI for this step~\cite{SASSIFI_ISPASS},
which injects errors into destination register of the selected instruction
based on the chosen error model (e.g., single- or double-bit flip) and bitmask.
Only one error is injected per application run. 
After error injection, the application is executed to completion, unless a
crash or a hang is detected. The injection outcomes are categorized based on
the exit status of the application, hang detection, error messages printed
during execution, differences in {\em stdout/stderr}, and program output
comparison from that of the error-free run.  Crashes and hangs are categorized
as {\it Arch DUEs}, failure symptoms as {\it Potential Arch DUEs}, and stdout
or program output differences as {\it SDCs}. Runs with same output as the
expected error free output and no error symptoms are categorized as {\it
Masked}. 

\section{Composing IPA and APA}
\label{sec:composition}

We compute SDC FIT rate estimate by composing results from IPA and APA
according to Equations~\ref{eqSDC} for each of our workloads. We call this
estimate {\it SDC TL-FIT} (TL is abbreviation for Two-Level).  We use issue IPC
averaged across all the kernels in an application as the scaling factor ($s$)
in the equation. We obtain it from {\it issued\_ipc} metric from {\it
nvprof}~\cite{NVPROF:Online}.

While we cover most of the dynamic instructions (80\% on average as shown in
Section~\ref{sec:apa-results}), we do not model SDC contribution from the
remaining instruction types (not shown in
Figure~\ref{fig:instruction_fraction}). Accounting for such instructions'
implementation-level and architecture-level propagation will increase the SDC
TL-FIT. 

We compare the TL-FIT with IPA- and APA-only estimates.  The fraction of
dynamic instructions covered in IPA-only and APA-only evaluations are same as
that covered in our TL-SDC calculations.  

{\bf Comparison with IPA-only:} 
For the IPA-only approach, we assume
that all arch\-i\-tec\-ture-le\-vel manifestations that corrupt instruction
output registers will result in application-level SDCs. We obtain IPA-only
results by using the same equation that we used to obtain TL-FIT, except that
we set $pSDC_{mn} = 1$ (in Equation~\ref{eqSDC}). 

{\bf Comparison with APA-only: } 
For the APA-only approach, we perform uniform arch\-i\-tec\-ture-le\-vel injections,
which assumes that the particle strikes in unprotected low-level state manifest
as single-bit flips in destination registers of executing instructions
uniformly similar to several prior studies~\cite{feng2010shoestring,
li2012classifying, fang2014GPU-Qin}.  We obtain APA-only results by setting the
values in the fourth column (single thread, single register, single bit) in
Table~\ref{tab:fbits-manifestation} to one and zero out all other values after
column two, except for FFMA and BRA. We set the fifth column value to one for
FFMA. We assume that errors in BRA instructions will result in DUEs and set the
third column value to one.  The APA-only approach estimates the probability of
SDCs per application, not the FIT rate.  So we compare the normalized APA-only
results with the TL-FIT rates mainly for the relative trends. 

{\bf Comparison with Beam Tests: }
To evaluate the accuracy of our model, we beam tested four workloads to compare
the SDC FIT rates with our TL-FIT rates.  We tested heartwall, lavaMD, CoMD,
and Lulesh. Details about the experimentation procedure is described in
Section~\ref{sec:ipa:beam}. For the experiments, we increase the input
size such that the GPU kernel execution is the significant portion of the total
execution time for
CoMD, Lulesh, and heartwall. We consider the kernel time as well as the time
spent in copying memory towards effective execution time.  For lavaMD the
fraction of time spent in copying memory is significant.  In these experiments,
we consider any indication of a failure as a DUE\@. For example, if we observe
a non-zero exit status or an error message in the system log, which we observe
using {\it dmesg}\/ utility, we categorize it as a DUE.  In these experiments,
whenever the program output file differs from the fault-free copy and there is
no indication of a failure,
we categorize the run as an SDC.

\section{Results}
\label{sec:results}

In this section, we quantify how low-level errors manifest at the
architecture-level (IPA) and the probability with which such manifestations
propagate to the application output (APA). We show a method to compose these two
results to estimate application FIT rates and compare the results with IPA- and
APA-only methods, as well as to end-to-end beam tests.  

\subsection{IPA}
\label{sec:ipa-results}

Table~\ref{tab:microbench-manifestation} shows the relative FIT rates for
different microbenchmarks (in the second column). We normalize the results to
the FIT rate of IADD microbenchmark.  We show the different events observed
during beam tests (e.g., crashes, hangs, architectural bit-flips) and their
relative rates in columns 3-5.  Values in these columns sum up to the value in
column 2 per row, if there is no rounding error.  

As explained in Section~\ref{sec:ipa:attribution}, we attribute observed events
to the four types of low-level bits. 
We show the attribution also in the Table (in the top row).  In the third
column, we show failures including hangs and crashes that happened before GPU
kernel started and during the kernel execution.  For some of these events, we
were not able to attribute to any specific type of bit. However, the majority
of these events were due to crashes/hangs before the GPU kernel execution and
we attribute them to IM-bits. Among all the events that were attributed to F-
and IS-bits, majority of the were architectural manifestations (as bit-flips in
instruction outputs) and many were due to F-bits. The observation that majority
of the architectural manifestations are attributed to F-bits suggests the
proposed two-level model can be accurate for application SDC rate estimation.

\begin{table}[tbp]
\begin{center}
    \scriptsize
    \caption{The architecture-level error manifestation rate per microbenchmark 
    from our beam testing campaign. We normalize the results to 
    the total FIT rate of the IADD microbenchmark.} 
    \vspace{-0.05in}
    \label{tab:microbench-manifestation}
	\begin{tabular}{|c|c|c|>{\centering\arraybackslash}p{1.6cm}|c|}
    \hline
		Bit Type & \multicolumn{2}{c|}{Any} & F & IS \\
    \hline

    \multirow{3}{1.2cm}{\centering Micro-benchmark} & 
	    \multirow{3}{1cm}{\centering Relative FIT} & 
	    \multirow{3}{1cm}{\centering Hangs \& Crashes} & 
		{\centering Architectural bit-flips \& crashes} & 
	    \multirow{3}{1.6cm}{\centering Architectural bit-flips} \\
    \hline

    IADD & 1 & 0.71 & 0.29 & \\ \hline

    FADD & 0.8 & 0.63 & 0.17 & \\ \hline

    IMAD & 0.72 & 0.47 & 0.12 & 0.14 \\ \hline

    FFMA & 0.98 & 0.69 & 0.21 & 0.08 \\ \hline

    LDS & 0.19 & 0.14 & 0.04 & 0.01 \\ \hline

    ISETP & 0.25 & 0.13 & 0.13 & \\ \hline

    BRA & 0.19 & 0.17 & 0.02 & \\ \hline

    \end{tabular}
\end{center}
\vspace{-0.1in}
\end{table}

\begin{table}[tbp]
    \scriptsize
    \caption{The architecture-level manifestation rate per issued
	    instruction for the F-bits. We scale the FIT rates attributed to
	    F-bits in Table~\ref{tab:microbench-manifestation} with the target
	    instruction's issue rate to obtain FIT rate per issued instruction.
	    These results are normalized to the per issued instruction FIT rate of IMAD.} 
    \vspace{-0.05in}
    \label{tab:fbits-manifestation}
    \centering

    \begin{tabular}{|c|c|c|c|c|c|>{\centering\arraybackslash}p{1.1cm} |}
    \hline

    \multirow{6}{0.85cm}{\centering Instruc\-tion} & \multirow{6}{0.9cm}{\centering FIT per instruc\-tion per SM} & \multirow{6}{0.85cm}{\centering Crashes} &
    	\multicolumn{4}{c|}{Architectural bit-flips} \\ \cline{4-7}

    & & &
	    \multicolumn{3}{c|}{\multirow{4}{2.7cm}{\centering Single register in single thread}} & 
    1 reg per thread in 2 threads\\ \cline{4-7}
  
    & & &
	1 bit & 2 bits & 3+ bits & 3+ bits \\ \hline

    IADD & 0.7 & & 0.54 & 0.08 & 0.08 & \\ \hline

    FADD & 0.42 & & 0.08 & 0.17 & 0.17 & \\ \hline

    IMAD & 1 & & 0.40 & 0.40 & 0.20 & \\ \hline

    FFMA & 0.69 & & & 0.26 & 0.35 & 0.09 \\ \hline

    LDS & 0.40 & 0.20 & 0.20 & & & \\ \hline

    ISETP & 0.67 & & 0.67 & & & \\ \hline

    BRA & 0.21 & 0.21 & & & & \\ \hline

    \end{tabular}
\vspace{-0.1in}
\end{table}

We next study the distribution of the events attributed to F-bits. We show the
architectural manifestation rate per issued instruction in
Table~\ref{tab:fbits-manifestation}. We scale the FIT rates presented in the
fourth column in Table~\ref{tab:microbench-manifestation} based on the
instruction issue rate of the target instruction to obtain the FIT rate per
issued instruction, shown in the second column in
Table~\ref{tab:fbits-manifestation}. The rates of different instruction
output-level manifestations are shown in the subsequent columns.  All the
values after column 2\ in each row sum up to the value in column 2 if
there is no rounding error.  These manifestations are (1)~crashes, (2)~single bit-flips,
(3)~double bit-flips, and (4)~3+ bit-flips in a single register in a single thread, and
(5)~3+ bit-flips in a same architectural register in two threads.  We later use the absolute
FIT rates (not shown here) while composing IPA and APA results. 
Based on these beam test results, we draw the following conclusions. 

\begin{itemize}[noitemsep, leftmargin=*]

	\item Most of the hangs and crashes in our compute-heavy microbenchmarks are caused
	by the I-bits, which indicate that modeling F-bits alone is insufficient
	to accurately estimate the DUE rates even for the compute-heavy workloads.  

	\item Most of  manifestations that flip bits at instruction output-level are
	attributed to the F-bits, and these manifestations corrupt just one register in
	one or two threads. (1) This simplifies modeling such manifestations and quantifying their 
	application-level propagation probabilities using APA.  (2) This also 
	implies that modeling F-bits alone for compute-heavy workloads can provide a
	reasonable SDC FIT rate estimate.

\end{itemize}

\subsection{APA}
\label{sec:apa-results}

\begin{figure}[tbp]
    \centering
    \includegraphics[width=0.5\textwidth]{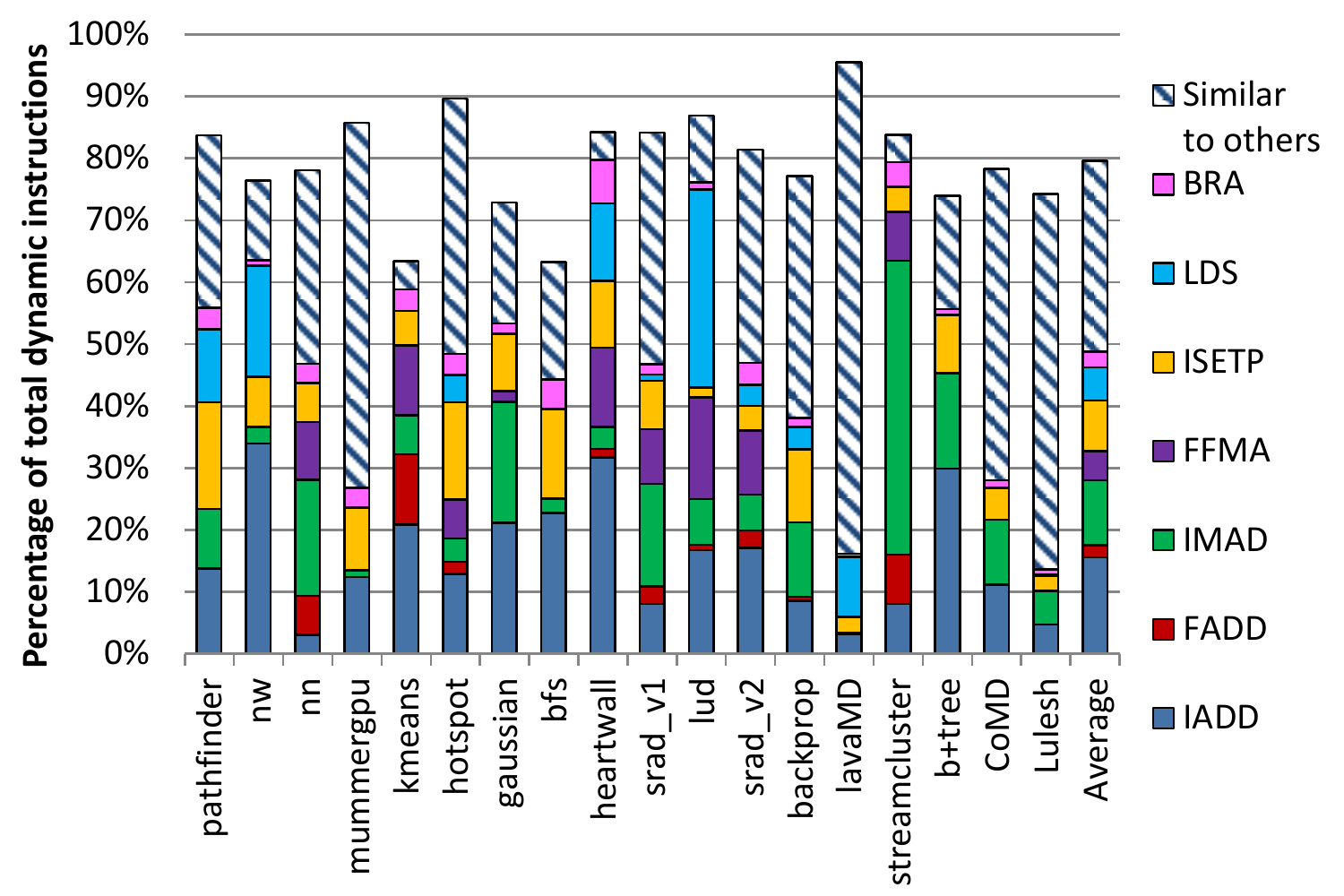}
    \vspace{-0.25in}
    \caption{The fraction of different instruction types in our workloads.}
    \label{fig:instruction_fraction}
    \vspace{-0.1in}
\end{figure}

{\bf Instruction Distribution:}
We plot the percentage of total dynamic instructions for IADD, FADD, IMAD,
FFMA, LDS, ISETP, and BRA in Figure~\ref{fig:instruction_fraction}. The results
show that the seven instructions represent up to 80\% of the dynamic
instructions (with an average of 49\%) for the workloads. We also plot the
percentage of dynamic instructions that we group together according to the
description in Section~\ref{sec:apa}. The grouping increases the representation
to 80\% on average for our workloads. 

\begin{figure}[tbp]
    \centering
    \includegraphics[width=0.5\textwidth]{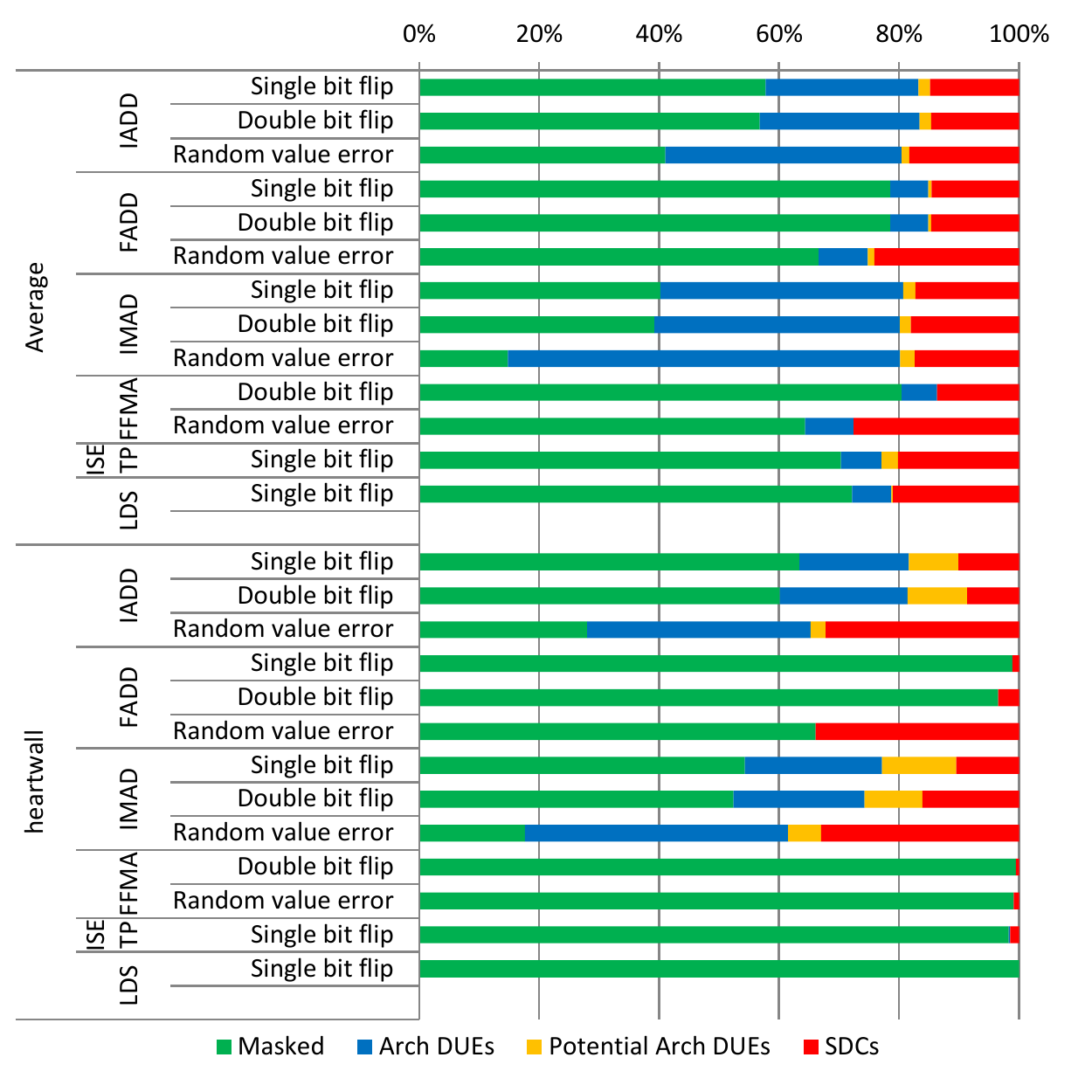}
    \vspace{-0.25in}
    \caption{Masking, DUE, and SDC derating factors for different error models
    and opcode groups. Two results are shown: those averaged across all of our
workloads and the detailed numbers for one application (heartwall).} 
    \label{fig:sassifi_summary}
    \vspace{-0.1in}
\end{figure}

{\bf Architecture-level Error Injections: }
We conduct SASSIFI error injection campaigns for CoMD, Lulesh, and all
workloads from the Rodinia benchmark suite using the error model based on the
architectural manifestations observed in the IPA experiments.  Specifically, we
inject errors per instruction type based on
Table~\ref{tab:fbits-manifestation}. For the 3+ bit flips in a register (from
Table~\ref{tab:fbits-manifestation}) in a single register, we inject a random
value in the destination register. We do not inject errors that corrupt two
threads.  We perform over 115,000 error injection runs, taking approximately
250 hours on a Tesla K20 GPU. Note that these results do not change with the
GPU, as long as the target ISA version (e.g., sm\_35) is fixed. A summary of
the results averaged across all the workloads is shown in
Figure~\ref{fig:sassifi_summary}. We do not inject errors in branch
instructions because we observed no events that manifest as bit-flips at the
architecture-level state. 

This figure, as well as the detailed per-application results (we only show
heartwall here for brevity) show that injecting single and double bit flips
into destination registers results in similar outcomes.  We note that
injecting random value errors, however, often produces different outcome
distribution.  Hence, it is important to know what proportion of particle
strikes that manifest as 3+ bit flips in the instruction outputs. 
The detailed per application results show that the errors in destination
registers in different instruction types behave differently. 
Based on these results, we draw the following conclusions, which suggest that
APA is a necessary step in estimating application SDCs. 

\begin{itemize}[noitemsep, leftmargin=*]
		
\item Propagation of architecture-level error depends heavily on the application, necessitating application-specific analysis.
\item Error propagation probabilities depend heavily on instruction type within an application.
\item Error propagation probabilities depend on bit-flip model for a given instruction type. 

\end{itemize}
 
\subsection{SDC Rates When Composing IPA and APA}
\label{sec:fit-results}

We compute SDC TL-FIT rate by composing results from IPA and APA
as described in Section~\ref{sec:composition} and Equations~\ref{eqSDC}. 
We show issue IPCs (scaling factor) in
Table~\ref{tab:scaling_factor}.  High issue IPC for a workload implies that
the it is able to keep the GPU pipelines busy and a high potential for high TL-SDC.  

\begin{table}[tbp]
	\caption{Average instruction issue rate for our workloads.} 
\vspace{-0.05in}
\label{tab:scaling_factor}
\scriptsize 
\centering
\begin{tabular}{|>{\centering\arraybackslash}p{1.4cm}|>{\centering\arraybackslash}p{0.6cm}|>{\centering\arraybackslash}p{1.2cm}|>{\centering\arraybackslash}p{0.6cm}|>{\centering\arraybackslash}p{1.5cm}|>{\centering\arraybackslash}p{0.6cm}|}
\hline
\multirow{2}{*}{\centering Workload} & Issue IPC &
\multirow{2}{*}{\centering Workload} & Issue IPC &
\multirow{2}{*}{\centering Workload} & Issue IPC \\
\hline
pathfinder   &  2.45 &
nw           &  0.51 &
nn           &  1.26 \\
mummergpu    &  1.76 & 
kmeans       &  0.78 &
hotspot      &  3.49 \\
gaussian     &  0.75 &
bfs          &  1.27 &
heartwall    &  1.88 \\
srad\_v1     &  1.69 & 
lud          &  0.13 &
srad\_v2     &  2.61 \\
backprop     &  1.56 &
lavaMD       &  3.85 & 
streamcluster&  0.75 \\
b+tree		&  1.40 &
CoMD		&  2.81 &
Lulesh		&  1.62 \\
\hline 
\end{tabular}
\vspace{-0.1in}
\end{table}

\begin{figure}[tbp]
    \centering
    \includegraphics[width=0.5\textwidth]{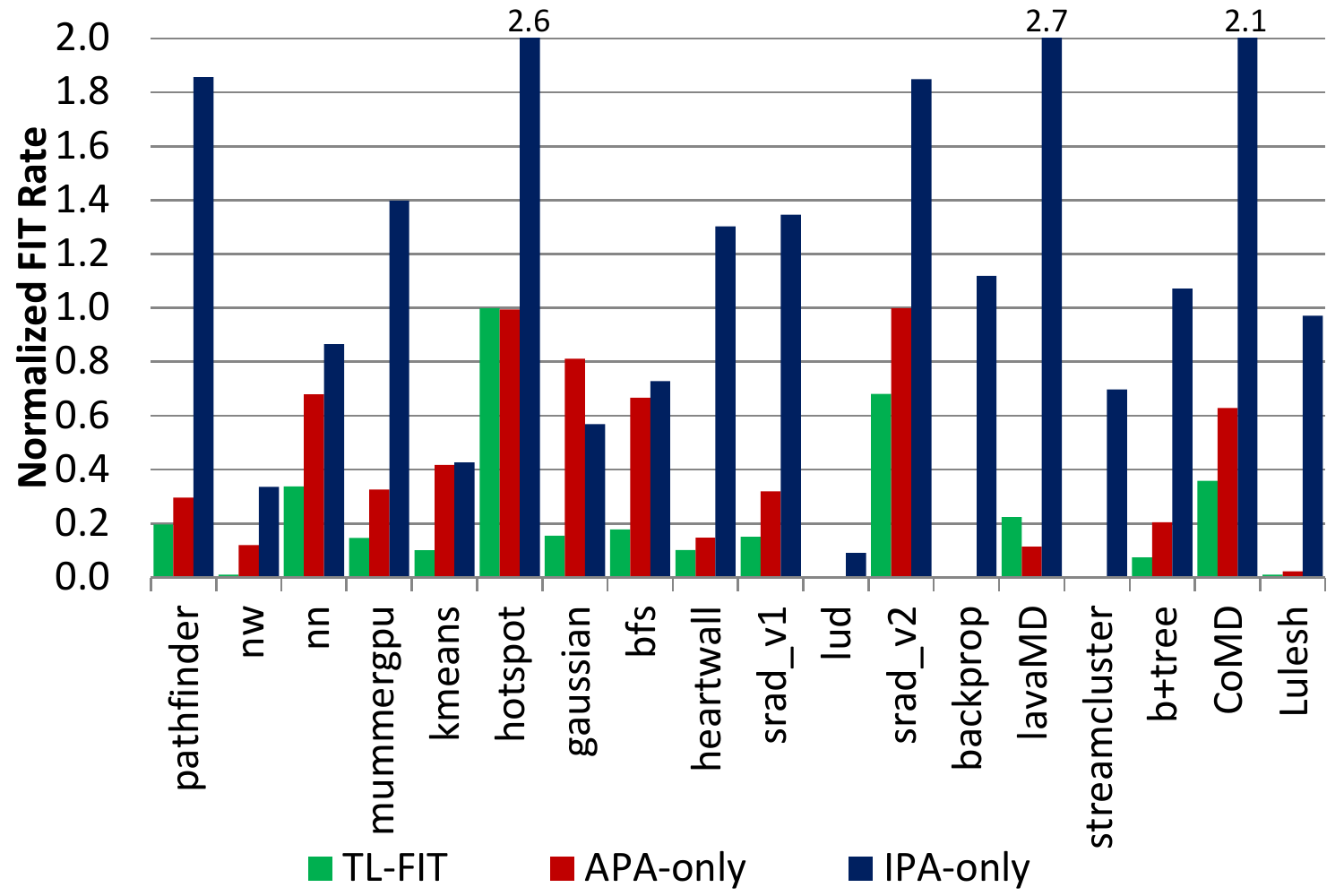}
    \vspace{-0.2in}
    \caption{Relative SDC FIT rate estimates (TL-FIT) for the F-bits 
		obtained using our compositional model.  The results are
		normalized to the SDC TL-FIT rate of hotspot. We also show
		APA-only SDC probabilities and IPA-only SDC FIT rates here for
		comparison, which are normalized to the APA-only SDC probability
		of srad\_v2 and the TL-FIT of hotspot, respectively.} 
    \label{fig:relative_fit}
\vspace{-0.1in}
\end{figure}

SDC TL-FIT rates computed using our method are shown in
Figure~\ref{fig:relative_fit} for different applications, normalized to the SDC
TL-FIT rate of hotspot.  We observe that either low APA SDC probabilities or
issue IPC implies that the SDC TL-FIT will be low. Results show that lud,
backprop, streamcluster, and Lulesh have low SDC TL-FIT rates and they all have
low APA SDC probabilities.  Hotspot and srad\_v2 are the two applications with
the highest SDC TL-FIT rates and they both have high APA SDC probabilities and
average issue IPC. 
%
We next compare our results with two approaches that either do not consider IPA
or APA and explain the differences. We call them APA-only and IPA-only,
respectively. 

{\bf Comparison with IPA-only: } 
IPA-only approach conservatively assumes that all arch\-i\-tec\-ture-le\-vel
manifestations will result in application-level SDCs. We
show the results in Figure~\ref{fig:relative_fit}, which are normalized to the
TL-FIT of hotspot.  As expected, the results show that the IPA-only FIT rates
are always higher than TL-FIT rates. They are, in fact, $>$10x higher for six
of our workloads.  The IPA-only results also show significantly differences in
the trends compared to TL-FIT. For the workloads with low APA-only SDC
probabilities, the difference between the TL-FIT and IPA-only FIT rates are
higher.  These results highlight the importance of performing IPA to quantify
the architecture-level error manifestation rates and APA to quantify the
application-level error propagation probabilities.

{\bf Comparison with APA-only:} 
The APA-only approach performs uniform architecture-level injections, similar
to many prior studies~\cite{feng2010shoestring, li2012classifying,
fang2014GPU-Qin}. It estimates the probability of SDCs per application given an
architecture-level error, not the FIT rate, making it difficulty to directly
compare with SDC TL-FIT.  We however compare the normalized SDC probabilities
in Figure~\ref{fig:relative_fit}. We normalize the applications' SDC
probabilities to the highest observed SDC probability (0.43 for srad\_v2).  

The results show significant differences in the trends compared to TL-FIT. For
example, the top two SDC-vulnerable applications according to APA-only results
have similar SDC probabilities, whereas the TL-FIT rates show a difference of
nearly 1.5x. Two of the top five SDC-vulnerable applications are different when
selected using the two evaluation techniques.  Primary reasons for these
differences are the following. (1)~The TL-FIT uses absolute rate of
archi\-tecture-level manifestations and does not rely on the accuracy of DUE
estimates. The APA-only approach, however, requires the relative DUE
probabilities to be accurate to estimate relative SDC probabilities. (2) The
APA-only approach does not consider the relative vulnerabilities of different
instructions and rates of different architecture-level manifestations. 
Our findings are inline with one of the prior studies (conducted on CPUs) which
also showed that high-level error propagation analysis alone may be highly
inaccurate when compared to studies that model errors at low-level using FPGA-
or RTL-based simulators~\cite{Cho2013DAC}. 

\begin{figure}[tbp]
   \centering
   \includegraphics[width=0.4\textwidth]{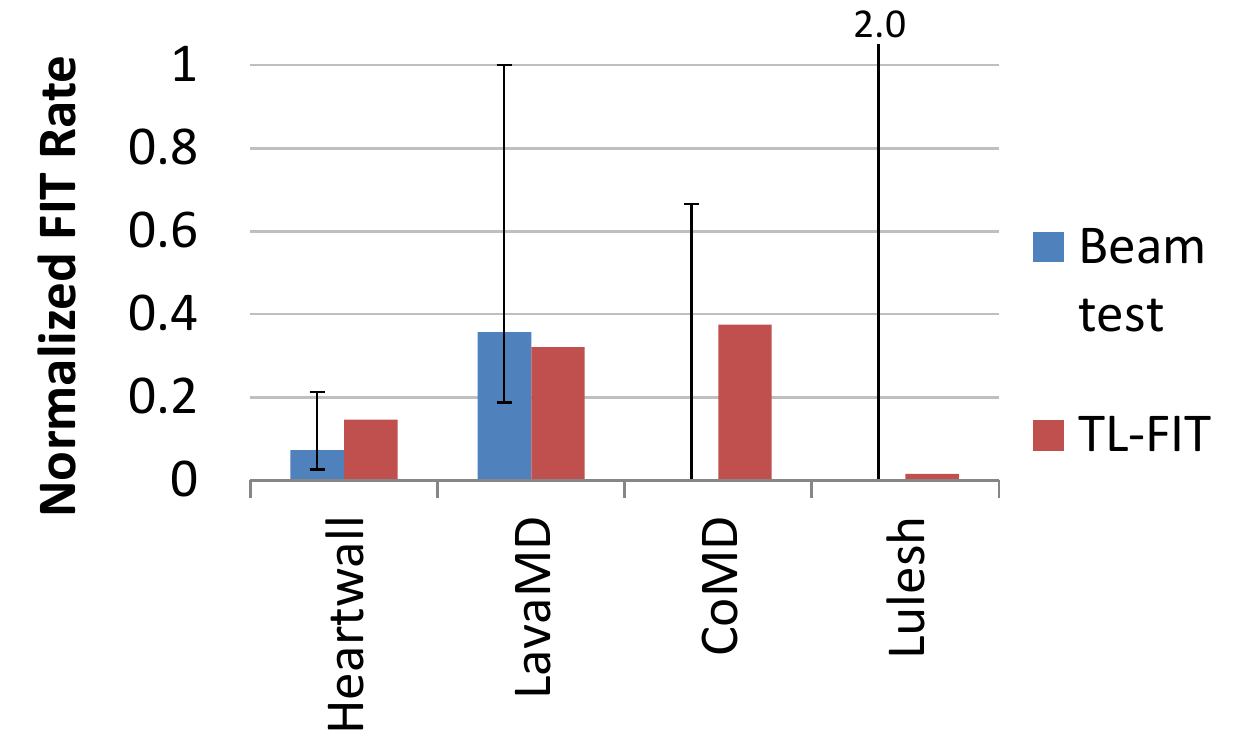}
   \vspace{-0.1in}
   \caption{Comparing application SDC FIT rates obtained from direct beam tests
	   with our TL-FIT rate estimates. SDC rates are normalized to the 95\%
   	   confidence upper limit of Lulesh's SDC FIT rates.} 
   \vspace{-0.1in}
   \label{fig:validation}
\end{figure}

{\bf Comparison with Beam Tests: } 
We compare SDC TL-FIT with the SDC FIT rate obtained by beam testing four
workloads. The results are shown in Figure~\ref{fig:validation}, which are
normalized to the 95\% confidence upper limit of the SDC rate of lavaMD.
Results show that the SDC TL-FIT rates are close to the beam test results and
are within the confidence intervals. While the error bar from beam test measurement
for Lulesh is high, we expect the SDC FIT to be low (close to 0)
based on several prior studies which show that only a small fraction of
low-level errors propagate to the Lulesh's output~\cite{Hamartia, Li2018,
Fang2016}. Our APA results also show that only a small percentage of
architectural manifestations propagate to the output, which also suggests that the SDC
FIT for this application should be small compared to other workloads.  A prior
study showed that CoMD is significantly more susceptible than
Lulesh~\cite{Hamartia}. Our SDC TL-FIT results also show the same trend.



\section{Discussion}
\label{sec:discussion}
\vspace{-0.1in}

\subsection{Applications}

{\bf Understanding Application Resilience: }
Identifying application characteristics that correlate well with either SDCs or
DUEs is an interesting application. For example, certain programming constructs
or design patterns may be associated with greater error propagation.
Identification of those constructs or patterns would allow programmers to
adjust their application code for higher reliability. This would also aid the
development of libraries that have higher resilience, which can be substituted
based on the error rate and overhead targets. 

We inspected CUDA and SASS code of two workloads with highest SDC TL-FIT
(hotspot and srad\_v2), and one with least SDC TL-FIT (lud).  For hotspot and
srad\_v2, only a small fraction of SASS instructions are used for control and
address computation, whereas the fraction of instructions used for address
computation is high for lud. These observations along with the
results presented in Figure~\ref{fig:relative_fit} indicate that whenever the
fraction of instructions used for control and address computation is high, the
likelihood of DUEs is also high in APA.  Interestingly, prior studies have also
made similar observations~\cite{Ramachandran2011SELSE,
feng2010shoestring}. 
Address randomization can be used to increase the likelihood of
DUEs in the APA step. 

{\bf Deciding Which Workloads Need Protection: }
The presented approach can be directly employed to decide which workloads need
protection. For example, for an HPC system with large number of GPUs and a
mean-time-to-SDC target that is usually in weeks or months, a per GPU target 
SDC FIT rate can be derived. This limit can be placed on
Figure~\ref{fig:relative_fit} to select applications that exceed the SDC
target. 
As an example, with the per-GPU SDC target of $0.5$ in
Figure~\ref{fig:relative_fit}, we would select hotspot and srad\_v2 for protection.

Our approach can also be used to identify highly vulnerable kernels within an
application to enable selective software-implemented error detection and
correction methods~\cite{Yim2011Hauberk, Wadden2014Redundancy,
Jeon2012WarpedDMR} for cost-effective protection.

\subsection{Accounting for bit-flips in IS- and V-bits} 
\label{sec:discussion:vbits}

Our IPA results (Section~\ref{sec:ipa-results}) show that a small fraction of
architecture-level bit-flip manifestations are attributed to IS-bits. These
manifestations corrupt registers in multiple threads in one or two warps
(categories 5-8 in Section~\ref{sec:ipa:attribution}). By definition, the rate of
such manifestations is device- and application dependent. Predicting these
rates using existing performance counters is an interesting research direction.
Based on our experimental data, we find inverse correlations with two
performance metrics {\it stall\_inst\_fetch} and {\it
issue\_slot\_utilization}, as reported by $nvprof$~\cite{NVPROF:Online}. The
correlation coefficients are -0.97 and -0.58, respectively. Developing a model
to predict such manifestations would require more experiments and validation.
Our APA methodology is capable of injecting these architecture-level
manifestations.

The contribution of V-bits can become dominant for instructions that stress the
memory subsystem (global memory loads/stores). Bit-flips in unprotected
structures in the memory subsystem can propagate as bit-flips into a cache or
memory line. Such a errors can corrupt a single or multiple instructions
depending on the reuse characteristic of the line. 
We conducted IPA (beam) experiments using two microbenchmarks of load
instruction that always hit and miss in L2 cache, respectively, stressing the
memory subsystem in different ways.  We observed higher crash/hang rate
compared to the tests that focus on exercising F-bits. Interestingly, we only
observed single bit flips in the destination registers of one instruction.
Since the memory subsystem of GPUs has several layers of buffers, more
experiments are needed to accurately model contribution of V-bits. 
A simulator that models some of the major unprotected buffers in the memory
subsystem (e.g., SASSIFI with a detailed cache model) may be better suited to
perform IPA.  

\subsection{Other Considerations} 

Our approach, as presented in the paper, is a post-silicon estimation technique
because IPA and APA experiments are performed on production devices. This
approach can potentially be used to estimate FIT rates of future generation
GPUs by (1)~scaling and extrapolating IPA results based on technology and
architectural changes and (2)~replacing the IPA and APA experiments with
low-level pre-silicon simulation-based approaches.
A prior study employed Register Transfer- and Gate-level error simulations in
an out-of-order processor to quantify how low-level errors propagate to the
instruction-level~\cite{Maniatakos2011}. 
While this method provides high control in performing IPA, the paper focused on
limited microarchitecture-level units (e.g., the scheduler and ROB). Employing
such a method to more low-level units, as permitted by the available
engineering resources, can be used to perform pre-silicon IPA. 

We do not consider multiple-bit faults induced by particle strikes in ECC
protected SRAM structures in our model because they can be converted into
correctable events with an appropriate level of bit-interleaving.  

Several field studies have been performed on large-scale HPC installations with
thousands of GPUs~\cite{Martino2014DSN, Tiwari2015HPCA}. These studies measure
the frequency of interrupts caused by GPU errors to estimate raw error and DUE
rates, but not the SDC rates. Such studies are complementary to the two-level
as we focus on SDC rate estimation.

\section{Conclusions}

This paper presents a new two-level methodology to estimate SDC rates of
applications running on production GPUs.  Our approach first quantifies how
particle strikes in low-level unprotected state manifest at the
architecture-level through accelerated beam experiments.  We then perform
near-silicon-speed architecture-level error injections to quantify how such
manifestations propagate to the program output.  Composing these two steps,
which we call IPA (implementation-level analysis) and APA (architecture-level
analysis), respectively, we estimated SDC rates for the workloads from Rodinia
benchmark suite and two DOE mini-apps. This two-level approach allows us to
estimate SDC rate of any application by performing just the APA, once the IPA
is performed. IPA needs to be performed just once per GPU generation. 

We compare our SDC rate estimates with accelerated beam test results for four
of our workloads and find them to be close.  We also compare our results with
two other approaches that either perform just the IPA or APA.  Results show
that ignoring either IPA or APA often overestimates SDC rates and show
significantly different trends---the composition of the two is needed for
accurate reliability modeling.


\bibliographystyle{ieeetr}
\bibliography{bibliography}


\end{document}